\documentclass{Interspeech2024}
\usepackage{subcaption}
\usepackage{multirow}
\usepackage{hyperref}
\usepackage{paralist}

\interspeechcameraready

\title{An End-to-End Approach for Chord-Conditioned Song Generation}

\name[affiliation={1,\dagger}]{Shuochen}{Gao}
\name[affiliation={1,\dagger}]{Shun}{Lei}
\name[affiliation={2}]{Fan}{Zhuo}
\name[affiliation={2}]{Hangyu}{Liu}
\name[affiliation={2}]{Feng}{Liu}
\name[affiliation={1}]{Boshi}{Tang}
\name[affiliation={1}]{Qiaochu}{Huang}
\name[affiliation={2,*}]{Shiyin}{Kang}
\name[affiliation={1,3,*}]{Zhiyong}{Wu}

\address{
  $^1$ Shenzhen International Graduate School, Tsinghua University, Shenzhen, China\\
  $^2$ Kunlun Skywork Technology Co., Beijing, China\\
  $^3$ Peng Cheng Lab, Shenzhen, China
  }
\email{\{gsc22,leis21\}@mails.tsinghua.edu.cn\thanks{$\dagger$Equal contribution.}, 
zywu@sz.tsinghua.edu.cn\thanks{*Corresponding author.}}


\begin{document}

\maketitle

\begin{abstract}
    The Song Generation task aims to synthesize music composed of vocals and accompaniment from given lyrics. While the existing method, Jukebox, has explored this task, its constrained control over the generations often leads to deficiency in music performance. To mitigate the issue, we introduce an important concept from music composition, namely \textit{chords}, to song generation networks. Chords form the foundation of accompaniment and provide vocal melody with associated harmony. Given the inaccuracy of automatic chord extractors, we devise a robust cross-attention mechanism augmented with dynamic weight sequence to integrate extracted chord information into song generations and reduce frame-level flaws, and propose a novel model termed Chord-Conditioned Song Generator (CSG) based on it. Experimental evidence demonstrates our proposed method outperforms other approaches in terms of musical performance and control precision of generated songs\footnote{Music sample: \href{https://thuhcsi.github.io/interspeech2024-CSG/}{https://thuhcsi.github.io/interspeech2024-CSG}}.

\end{abstract}

\section{Introduction}

Music, as a ubiquitous art form, plays a significant role in people's lives. Music that includes both accompaniment and vocals is referred to as songs, where well-designed songs necessitate a harmonious blend of vocals and accompaniment. The task of Song Generation, which synthesizes songs from lyrics, can play a critical role in the entertainment industry.

Early endeavors in music generation leveraged symbolic representations to produce score parameters \cite{ji2023survey}, which were then rendered into music. However, symbolic music is confined to fixed instrumental timbres and lacks expressiveness. Recent years have witnessed an emergence of End-to-End Music Generation models that generate musical audio through text prompts \cite{agostinelli2023musiclm,musicldm,huang2023noise2music,lam2024efficient} and melody control \cite{copet2024simple}. Yet, End-to-End Music Generation often struggles to produce meaningful vocals, frequently resulting in gibberish. Conversely, the Singing Voice Synthesis (SVS) field focuses on generating singing voices from lyrics and scores, with existing efforts \cite{zhou22f_interspeech,zhang2023wesinger,zhou2023bisinger,lei2023unisyn} capable of producing high-quality vocals. However, SVS-generated singing often lacks accompaniment and requires users to provide music scores, revealing a deficiency in the models' song composition capabilities.

To address the challenges faced by Music Generation and Singing Voice Synthesis, Jukebox \cite{dhariwal2020jukebox} introduces lyrics as a control condition on top of text prompts, enabling autonomous song generation.
However, it primarily models based on acoustic feature sequences, which impedes its ability to assimilate high-level music theory knowledge. Moreover, it exhibits limited control over the music generation process, often resulting in outputs that often lack musicality.

In this work, we present an innovative end-to-end Chord-Conditioned Song Generator (CSG), which introduces chord condition for generating condition-compliant songs. 
Chord, as an important concept of the song, forms the foundation of accompaniment and provides vocal melody with associated harmony \cite{briot2017deep}. Even simple chords can sustain the basic auditory sensation of accompaniment and, combined with vocals, create melodious songs. For instance, guitar playing and singing involve coupling guitar chords with matching vocals. Given the integral relationship chord shares with both accompaniment and vocals, it serves as a straightforward and effective control condition for generating both components.
To our knowledge, this is the first instance where chords have been used as a control condition in the domain of End-to-End Music Generation. Previously, only a portion of Symbolic Music Generation efforts utilized chord control \cite{li2023chord,li2023melodydiffusion,choi2021chord}.
Chord control simplifies manipulation significantly over melody and score control. Through CSG, even non-expert users, lacking formal musical theory knowledge, can employ standard chord progressions like `6451', `4536', or custom sequences, facilitating the creation of unique, harmonious songs.
Following \cite{audiolm,agostinelli2023musiclm}, CSG employs a Self-Supervised Learning (SSL) model to extract semantic tokens, serving as substitutes for acoustic features.
Moreover, given that existing methods for automatic chord extraction suffer from low precision issues, merely incorporating chords does not enable the model to effectively learn the relationship between chords and music, thereby impacting the musicality of the generated songs. To address this issue, we propose an innovative Attention with Dynamic Weights Sequence (DWS) that, while integrating chords with lyrics and songs, also assesses the correctness of chords frame by frame. By reducing interference from erroneous data and increasing the model's confidence in accurate chord data, this approach simultaneously enhances the musicality and control precision of the generated songs.

The main contributions of our paper are:
\begin{inparaenum}[(1)]
  \item We introduce chords as the control condition for song generation, effortlessly and efficiently enhancing the musicality of the generated songs.
  \item We propose an innovative Attention with DWS to improve the precision of control and the musical performance of the generated songs.
\end{inparaenum}



\section{Methodology}

\begin{figure*}[t]
  \begin{subfigure}[t]{0.5\textwidth}
      \centering
      \includegraphics[width=0.85\textwidth]{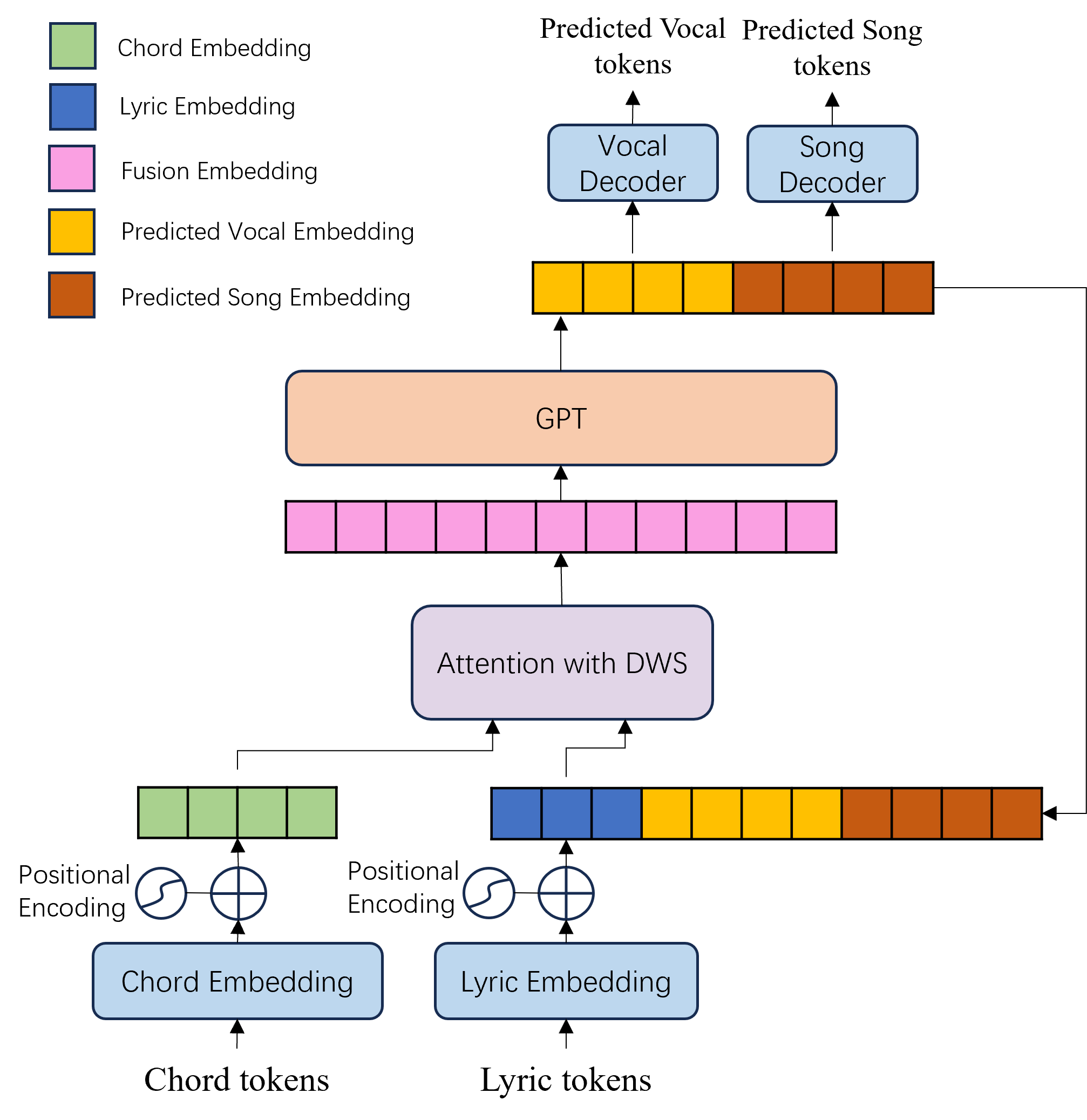}
      \caption{Structure of the Chord-Conditioned Song Generator (CSG)}
      \label{fig:1a}
  \end{subfigure}
  \hfill
      \begin{subfigure}[t]{0.5\textwidth}
      \centering
      \includegraphics[width=0.75\textwidth]{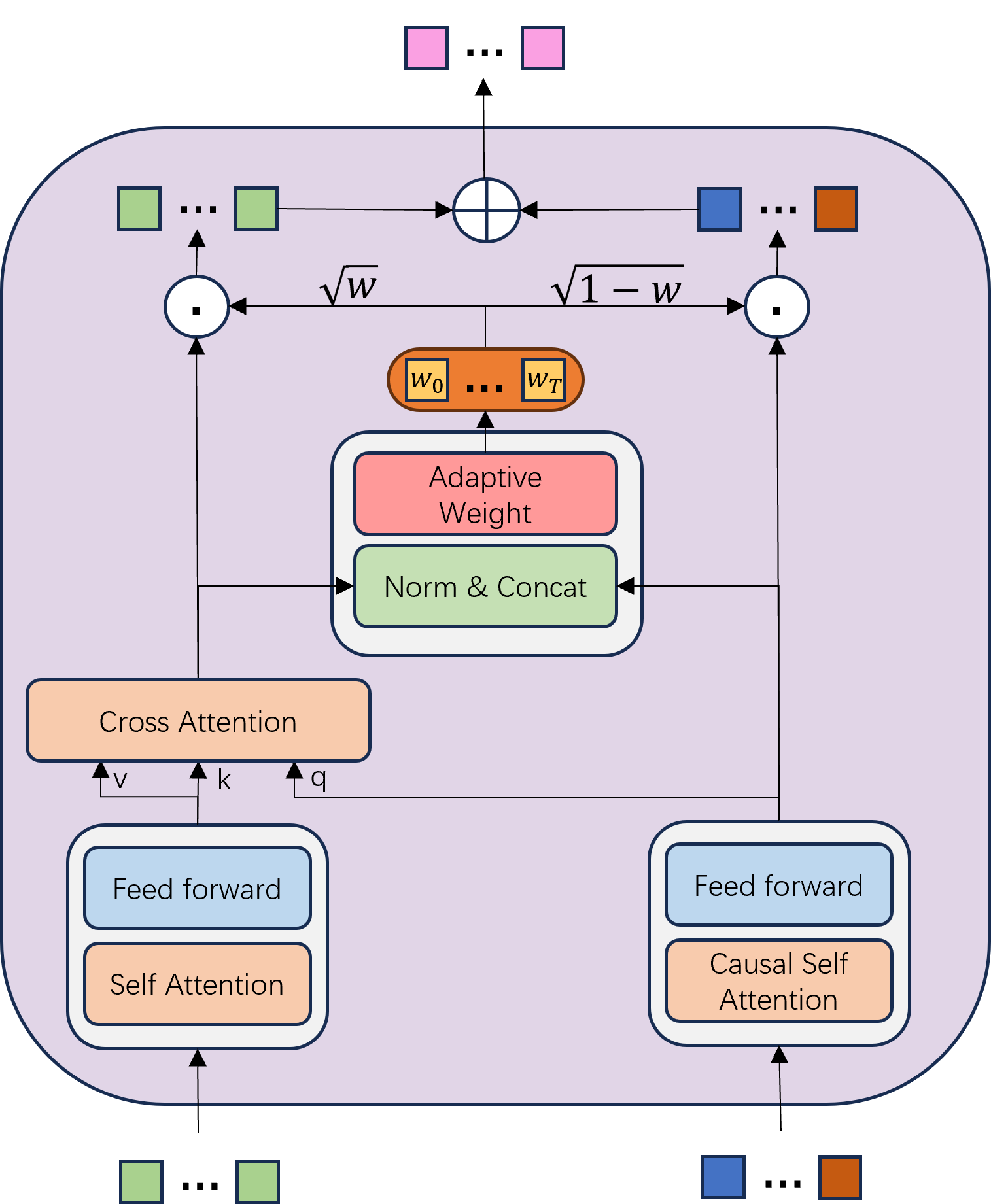}
      \caption{Details of Attention with Dynamic Weights Sequence (DWS)}
      \label{fig:1b}
  \end{subfigure}
  \caption{The overall architecture of CSG with proposed Attention with DWS}
  \label{fig:1}
\end{figure*}

As shown in Figure~\ref{fig:1a}, given chord and lyric tokens as inputs, CSG generates frames of songs in an autoregressive manner. First, we map the input tokens to high-dimensional space. 
Then the chord and lyric embeddings
are combined with vocal embeddings as well as song embeddings from previous frames, and fed into the Attention with DWS for semantics fusion. After that, the fusion embedding gets populated into a GPT module \cite{GPT} to generate vocal and song embeddings for the current frame, which are finally decoded by two decoders for token prediction.
At the training stage, the GPT and Attention with DWS get initialized from random weights and are trained together, while the lyric tokens are extracted by a tokenizer in pre-trained BERT\footnote{BERT: \href{https://github.com/google-research/bert/tree/master}{https://github.com/google-research/bert/tree/master}} \cite{devlin2018bert}. Vocal and song tokens are extracted by pre-trained BEST-RQ\footnote{BEST-RQ: \href{https://github.com/lucasnewman/best-rq-pytorch}{https://github.com/lucasnewman/best-rq-pytorch}} \cite{borsos2023soundstorm,chiu2022selfsupervised}. Chord token extraction will be explained in Section 2.1. During inference, we keep all the model weights fixed and get the chord\&lyric tokens from users. At the end of inference, a diffusion vocoder, adjusted based on Stable Audio\footnote{Stable Audio: \href{https://github.com/Stability-AI/stable-audio-tools}{https://github.com/Stability-AI/stable-audio-tools}} \cite{evans2024fast}, facilitates the restoration from tokens to audio.
The following sections elaborate on the modules.



\subsection{Chord Token Extraction}

From the song data, we separate the background music and employ Autochord\footnote{Autochord: \href{https://github.com/cjbayron/autochord}{https://github.com/cjbayron/autochord}} to extract chord progressions from it. The extracted chord progressions consist of three components: chord roots, interval relations, and durations. Specifically, there are twelve possible chord roots: \textit{C, C\#, D, D\#, E, F, F\#, G, G\#, A, A\#, B} and four possible interval relations: \textit{Major}, \textit{Minor}, \textit{Augmented}, and \textit{Diminished}. The 48 possible combinations derived from them can cover the vast majority of chords. We further quantified the chord progressions into sequences, analyzing them at intervals corresponding to a 50Hz rate, with each frame quantized into one of 48 possible tokens. These tokens, determined by their root note and interval relation combination, are encoded as integers ranging from 0 to 47.




\begin{table*}[t]
  \caption{Evaluation results of different song generation methods. The results of mean opinion scores (MOS) in user study are shown with 95\% confidence intervals. The last row shows the ablation study. ``w/o Attention with DWS (concatenation)'' means using the concatenation method, and ``w/o Attention with DWS (cross-attention)'' means using cross-attention without DWS.}
  \label{tab:1}
  \centering
  \begin{tabular}{ccccc}
    \toprule
    \multirow{2}{*}{\textbf{Systems}}  & \multirow{2}{*}{$\mathrm{\textbf{FAD}}_{vgg}\downarrow$} & \multirow{2}{*}{\textbf{SIM$\uparrow$}}  & \multicolumn{2}{c}{\textbf{User Study$\uparrow$}} \\
                                         &               &               &MOS (MP)                 &MOS (CA)   \\
    \midrule
    Jukebox                              & 14.06         & -             & $2.46 \pm 0.13$   & -  \\
    GPT-only                             & \textbf{6.80} & 0.09          & $3.12 \pm 0.09$   & $2.65 \pm 0.15$  \\
    Ours                                 & 7.35          & \textbf{0.61} & $\textbf{3.74} \pm \textbf{0.09}$                                                                             & $\textbf{3.91} \pm \textbf{0.14}$    \\

    \midrule
    w/o Attention with DWS (concatenation)    & 7.67          & 0.52          & $3.17 \pm 0.09$   & $3.40 \pm 0.14$  \\
    w/o Attention with DWS (cross-attention)  & 7.71          & 0.46          & $3.24 \pm 0.10$   & $3.05 \pm 0.15$  \\

    \bottomrule
  \end{tabular}
\end{table*}

\subsection{Attention with Dynamic Weights Sequence (DWS)}
To integrate chords with lyrics and songs, the simplest method is to concatenate the chord sequence with the audio sequence in the embedding dimension or to use cross-attention. However, concatenation allows the autoregressive prediction of the song to only see the preceding chord sequence, not the subsequent chords. Additionally, the measured chord extraction accuracy of the Autochord is 67.33\%, which means our chord data inevitably contains some noise or even errors. These inaccuracies can interfere with the model's learning, whether through concatenation or cross-attention. To address this issue and improve the precision of control, we propose Attention with Dynamic Weights Sequence (DWS).

As illustrated in Figure~\ref{fig:1b}, Attention with DWS employs a dual-path architecture in which each path incorporates causal or non-causal transformer blocks to facilitate temporal alignment learning within sequences. Subsequently, a cross-attention mechanism is harnessed to synchronize chord embeddings with lyric-audio embeddings, resulting in the alignment output $C$:
\begin{align}
  C &= softmax (\frac{Q_{lyric-audio}K_{chord}^T}{\sqrt{d_k}})V_{chord}
  \label{equation:eq1}
\end{align}

As a weighted average to chord embedding values, $C$ is inevitably affected by the inaccuracies in chord data. This means that $c_{t}$, representing elements of $C$ at specific time points, may undergo varying degrees of perturbation. To mitigate the impact of such inaccuracies, we introduce a temporally adaptive weighting network. This network is designed to assess the correlation between chord embeddings and audio embeddings sequentially, on a frame-by-frame basis, leading to the generation of the weight sequence $W$:
\begin{align}
  W &= \sigma (M([C;A]))
  \label{equation:eq2}
\end{align}

Here, $A$ denotes the lyric-audio embeddings computed by causal transformer blocks, and $M$ is a mapping function that calculates temporal weights for each frame. $\sigma$ is the sigmoid activation function providing normalization. The generated weight sequence $W$ then evaluates chord utilization across different moments. Consequently, the fusion embedding sequence $F$, which integrates both global chord information and preceding audio data, is represented as follows:
\begin{align}
  F &= \sqrt{W} \circ C+\sqrt{1-W} \circ A
  \label{equation:eq3}
\end{align}

Compared with concatenation and cross-attention, Attention with DWS enhances the robustness of the Fusion Embedding Sequence $F$ towards chords containing inaccuracies while having access to global chord information. By employing a dynamic weight sequence to discern between correct and incorrect chords frame by frame, Attention with DWS ensures that the presence of incorrect chords during training does not diminish confidence in the correct chords. Thus, this enhances the precision of chord-conditioned control over the model. Furthermore, by minimizing the interference from incorrect chords, Attention with DWS enables the model to more effectively learn the musical correlation between chords and songs, thereby improving the musicality of chord-controlled song generation.

\begin{figure*}[t]
  \begin{subfigure}[t]{0.5\textwidth}
      \flushright
      \includegraphics[width=0.9\textwidth]{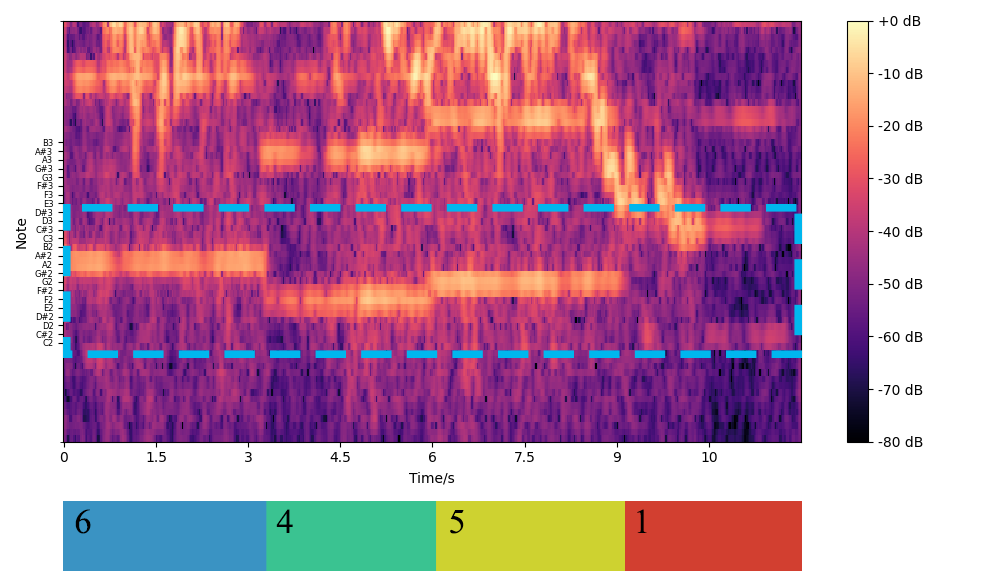}
      \caption{Proposed Model}
      \label{fig:2a}
  \end{subfigure}
  \hfill
      \begin{subfigure}[t]{0.5\textwidth}
      \flushright
      \includegraphics[width=0.9\textwidth]{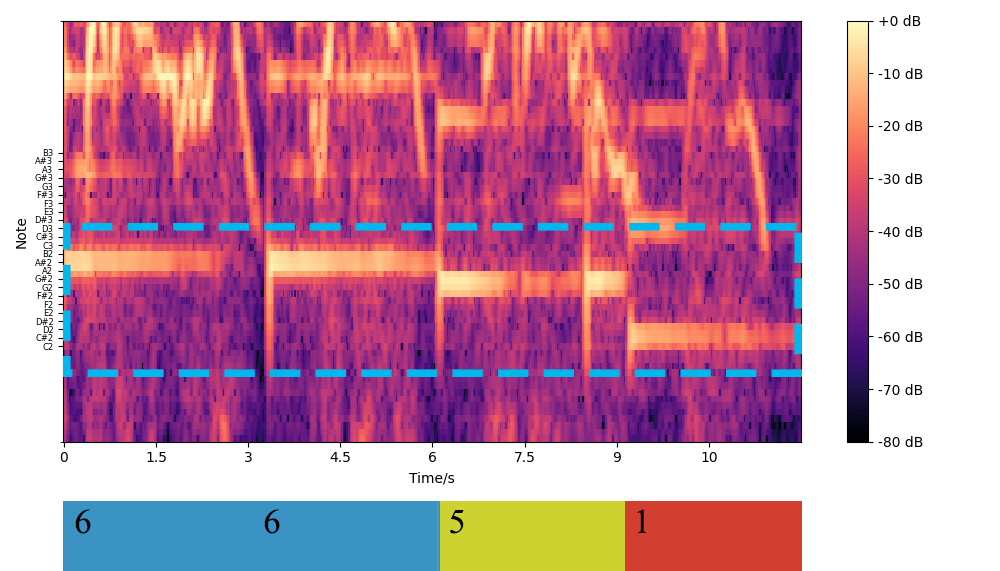}
      \caption{Concatenation}
      \label{fig:2b}
  \end{subfigure}
  \hfill
      \begin{subfigure}[t]{0.5\textwidth}
      \flushright
      \includegraphics[width=0.9\textwidth]{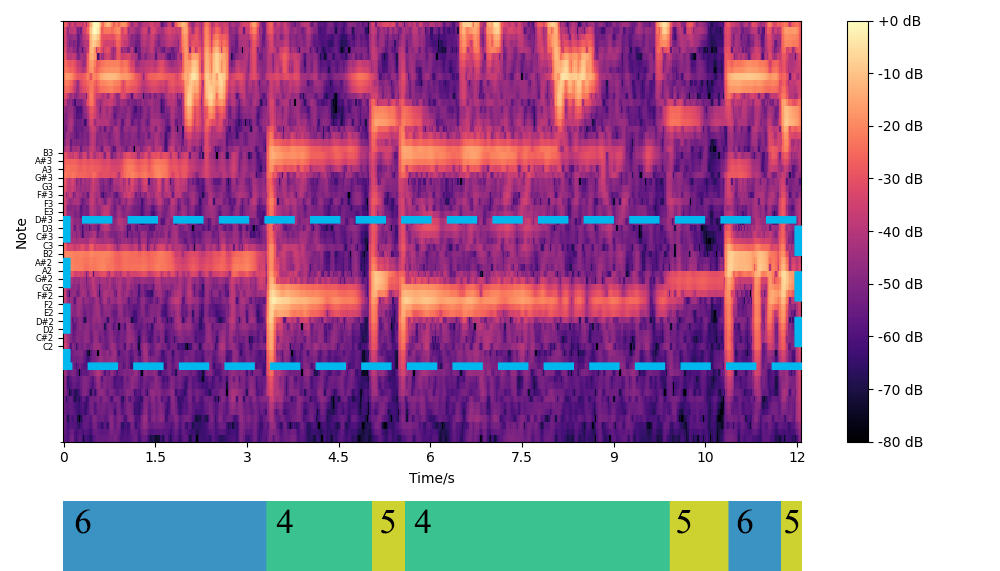}
      \caption{Cross-attention}
      \label{fig:2c}
  \end{subfigure}
  \hfill
      \begin{subfigure}[t]{0.5\textwidth}
      \flushright
      \includegraphics[width=0.9\textwidth]{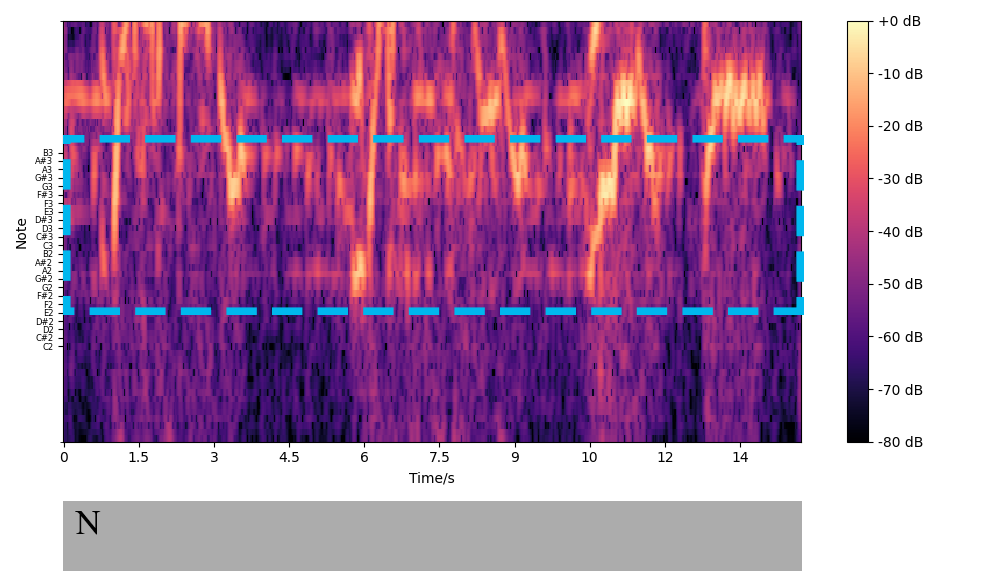}
      \caption{GPT-only}
      \label{fig:2d}
  \end{subfigure}
  \caption{Synthesized spectrograms with note labels. Blue rectangular frames and the color bars below the spectrograms highlight the chords of the generated songs.}
  \label{fig:2}
\end{figure*}

\section{Experiments}

\subsection{Dataset}
Given the lack of a large-scale, open-source lyrics-to-song dataset, we apply CSG to a proprietary dataset containing 554,467 English songs across various genres such as country, pop, rock, and rap. For preprocessing, the dataset is segmented into approximately 3 million vocal-only and accompaniment-only segments, each of which is annotated with the corresponding lyrics.
We randomly sample 5\% of the dataset and reserve it for validation and testing, while the rest is used for training.

\subsection{Experiment Setup}

We convert the chord, lyric, and audio tokens into 1024-dimensional embeddings. In Attention with DWS, both the self-attention block and the causal self-attention block consist of 2 layers, while the Adaptive Weight module is comprised of a single linear layer. The GPT model is constructed from 12 transformer blocks, each with a dropout rate of 0.1.
Training of the proposed model and ablation baseline models are conducted for 500,000 steps on seven NVIDIA GeForce RTX4090 GPUs, with a batch size of 4 per GPU, utilizing the Adam optimizer and a learning rate warm-up scheduler with a target learning rate of $8 \times 10^{-5}$ and a warm-up period of 32,000.

Jukebox is the sole work in the domain of song generation to date. We compare our model with the samples publicly available on the official Jukebox website.
Moreover, to demonstrate how incorporating a basic chord condition can markedly enhance musicality, we trained a GPT-only model without the chord condition as a deterministic \textbf{baseline} for comparison, employing the same training methodology as our proposed model.

\subsection{Results}

\subsubsection{Subjective Evaluation}
We employ two Mean Opinion Scores (MOS) to evaluate the capability of different models. 1) Musical Performance (MP): Evaluate the musicality of generated songs. 2) Chord Alignment (CA): Evaluate the correlation between the chords in the generated songs and the actual chords. 
For MP, we randomly select 15 sets of lyrics written by Jukebox researchers, which are not present in any existing dataset, to act as lyric inputs for each model. Additionally, we devise conventional chords, such as `6451', `4536', and `2516', as chord controls. Twenty-two participants are invited to rate these 15 song segments with identical lyrics, providing comprehensive scores for the songs' musicality.
For CA, six sets of chords, including both conventional (e.g., `6451', `4536', `2516', `1564') and unconventional (e.g., `1111', `1234') chords, are used to generate 12 samples. Twenty participants evaluate the correlation between the chords in the generated songs and the actual chords, considering auditory perception.

The last two columns of Table \ref{tab:1} report the results of the MOS evaluation.
Our proposed method achieves the highest MOS for both MP and CA. The MP of GPT-only music performance surpasses that of Jukebox, attributable to the modeling of semantic information by BEST-RQ, which improves the musicality of generated music. Moreover, the effective utilization of chord information is a key factor enabling our proposed model to outperform GPT-only in terms of musical performance. Owing to the GPT-only model's lack of chord conditioning, our proposed model significantly surpasses the GPT-only in terms of chord control precision.

\subsubsection{Objective Evaluation}

In terms of generation fidelity, we employ the Fréchet Audio Distance (FAD) \cite{kilgour2018fr} utilizing the VGGish model \cite{hershey2017cnn}, where a lower FAD indicates higher audio fidelity. We randomly extract 103 10-second song segments with unseen lyrics from the Jukebox website, and generate 1000 12-second song segments using unseen lyrics and chords by both GPT-only and CSG, respectively, to calculate FAD.

Furthermore, control precision serves as a critical metric for assessing the effectiveness of control. We specify random chords and pair them with unseen lyrics to generate 400 song segments. Utilizing Autochord, we extract the chords from these 400 segments and compare them with the specified chords to calculate the Similarity Index (SIM) for relative pitch accuracy, which calculates the proportion of the correct chords throughout the music. Meanwhile, we allow for a global shift in the key of the chords depending on the mode. Additionally, we compute the SIM for songs generated by the GPT-only model without chord conditions, serving as a reference baseline for uncontrolled generation.

The results for different methods are presented in Table \ref{tab:1}.
Both the GPT-only model and our proposed model exhibit lower FAD than Jukebox, attributable to the utilization of a vocoder based on Stable Audio, which enhances the fidelity of the generated audio. Besides, due to the introduction of chord control, our model exhibits a slight increase in FAD compared to GPT-only, while simultaneously demonstrating a significant improvement in SIM relative to conditions without chord control. The observed increase in FAD when incorporating control conditions is a typical phenomenon in music generation, and the improvement of SIM indicates that our model substantially enhances control over chords at the cost of a slight decrease in generation fidelity.

\subsubsection{Ablation Study}
We conduct ablation studies to further investigate the effectiveness of the attention mechanism with DWS. 
As shown in the last two rows of Table \ref{tab:1}, simple integrated methods including concatenation and cross-attention fail to identify inaccurate chords during training. This failure leads to a disruption in recognizing correct chords during inference, resulting in diminished control precision, as indicated by SIM and CA. Additionally, this failure also hampers the model's capacity to accurately determine the relationships between chords and music, leading to a reduction in the musicality of the generated songs, as evidenced by MP. The inaccuracy also slightly decreases the generation fidelity, as shown by FAD. 

\subsubsection{Case Study}
Beyond the quantitative metrics introduced earlier, the control precision of the proposed method can be directly demonstrated through qualitative analysis of spectrograms synthesized by various models under the same input conditions (Figure~\ref{fig:2}). When using chord condition as ``6451'', i.e., ``A:min-F:maj-G:maj-C:maj'', our proposed model generates a song with accurate chords, where the chord ``1'' concurrently resides in both notes ``C2'' and ``C3''. With the same chord and lyrics conditions, the concatenation model generates a song with chords ``6651'', and the cross-attention model generates a song with chords ``6454565''. Without using chords for control, controllable chord information is hard to be seen in the spectrogram.

\section{Conclusion}

In this work, we introduce a novel chord-conditioned song generation method, termed CSG, featuring our innovative Attention mechanism with DWS. This mechanism not only integrates chords with lyrics and songs but also reduces the impact of inaccurate chord data at the frame level. Experimental results demonstrate that CSG surpasses competing methods in musical performance through effective utilization of chord information, and Attention with DWS significantly enhances the musicality and control precision of generated songs.

\section{Acknowledgements}
This work is supported by National Natural Science Foundation of China (62076144), Shenzhen Science and Technology Program (WDZC20220816140515001, JCYJ20220818101014030) and the Major Key Project of PCL (PCL2022D01, PCL2023AS7-1).


\bibliographystyle{IEEEtran}
\bibliography{mybib}

\end{document}